\documentclass[a4paper,fleqn,usenatbib]{mnras}

\usepackage[T1]{fontenc}
\usepackage{ae,aecompl}

\usepackage{graphicx}	
\usepackage{amsmath}	
\usepackage{amssymb}	

\renewcommand {\deg}   {\mbox{$^\circ$}}

\newcommand   {\kms}   {\mbox{km\,s$^{-1}$}}
\renewcommand {\ga}    {\mbox{\rlap{\hbox{\lower5pt\hbox{$\sim$}}}\hbox{$>$}}}
\renewcommand {\la}    {\mbox{\rlap{\hbox{\lower5pt\hbox{$\sim$}}}\hbox{$<$}}}

\title[Origin of NRFs]{Cosmic-ray driven outflow from the galactic center\\
and the origin of magnetized  radio filaments}

\author[Yusef-Zadeh \& Wardle]{
F. Yusef-Zadeh$^1$\thanks{E-mail: zadeh@northwestern.edu} \&  M. Wardle$^2$
\\
$^{1}$CIERA, Department of Physics and Astronomy Northwestern University, Evanston, IL 60208\\
$^{2}$Dept of Physics and Astronomy,  Research Centre for Astronomy, Astrophysics\\
and Astrophotonics, Macquarie University, Sydney NSW 2109, Australia}

\date{Accepted XXX. Received YYY; in original form ZZZ}

\pubyear{2019}

\begin{document}
\label{firstpage}
\pagerange{\pageref{firstpage}--\pageref{lastpage}}
\maketitle

\def\msol{\hbox{$\hbox{M}_\odot$}}
\def\lsol{\hbox{$\hbox{L}_\odot$}}
\def\kms{km s$^{-1}$}
\def\Blos{B$_{\rm los}$}
\def\etal   {{\it et al.}}                     
\def\psec           {$.\negthinspace^{s}$}
\def\pasec          {$.\negthinspace^{\prime\prime}$}
\def\pdeg           {$.\kern-.25em ^{^\circ}$}
\def\degree{\ifmmode{^\circ} \else{$^\circ$}\fi}
\def\ut #1 #2 { \, \textrm{#1}^{#2}} 
\def\u #1 { \, \textrm{#1}}          
\def\nH {n_\mathrm{H}}
\def\ddeg   {\hbox{$.\!\!^\circ$}}              
\def\deg    {$^{\circ}$}                        
\def\le     {$\leq$}                            
\def\sec    {$^{\rm s}$}                        
\def\msol   {\hbox{$M_\odot$}}                  
\def\i      {\hbox{\it I}}                      
\def\v      {\hbox{\it V}}                      
\def\dasec  {\hbox{$.\!\!^{\prime\prime}$}}     
\def\asec   {$^{\prime\prime}$}                 
\def\dasec  {\hbox{$.\!\!^{\prime\prime}$}}     
\def\dsec   {\hbox{$.\!\!^{\rm s}$}}            
\def\min    {$^{\rm m}$}                        
\def\hour   {$^{\rm h}$}                        
\def\amin   {$^{\prime}$}                       
\def\lsol{\, \hbox{$\hbox{L}_\odot$}}
\def\sec    {$^{\rm s}$}                        
\def\etal   {{\it et al.}}                     
\def\la{\lower.4ex\hbox{$\;\buildrel <\over{\scriptstyle\sim}\;$}}
\def\ga{\lower.4ex\hbox{$\;\buildrel >\over{\scriptstyle\sim}\;$}}

\begin{abstract} 
Radio, X-ray and infrared observations of the inner few hundred pc of the Galactic
center have highlighted two characteristics to the ISM. The cosmic ray ionization
rate derived from molecular ions such as H$^+_3$, is at least two to three orders of
magnitudes higher than in the Galactic disk.  The other is bipolar X-ray and radio
emission away from the Galactic plane. These features are consistent with a scenario
in which high cosmic ray pressure drives large-scale winds away from the Galactic
plane. The interaction of such a wind with stellar wind bubbles may explain the
energetic nonthermal radio filaments found throughout the Galactic center. Some of
the implications of this scenario is the removal of gas driven by outflowing winds,
acting as a feedback to reduce the star formation rate in the central molecular zone
(CMZ), and the distortion of azimuthal magnetic field lines in the CMZ to vertical
direction away from the plane. The combined effects of the wind and vertical magnetic
field can explain why most magnetized filaments run perpendicular to the Galactic
plane. This proposed picture suggests our Milky Way nucleus has recently experienced
starburst or black hole activity, as recent radio and X-ray observations indicate.
\end{abstract}





\section{Introduction} 
MeerKAT observations of the Galactic center have recently discovered a
spectacular bubble of radio emission at 1.3 GHz extending over 430pc
\citep{heywood19}. This bubble covers prominent radio continuum
sources such as the radio arc at l$\sim0.2^\circ$, Sgr C at
l$\sim-0.5^\circ$, nonthermal radio filaments (NRFs), and the radio
lobes showing a mixture of warm ionized, dust  and synchrotron emission
\citep{sofue84,tsuboi95,bland-hawthorn03,zadeh04,law09,alves15,nagoshi19}. 
Radio recombination line (RRL) observations
of warm ionized gas in  the northern and southern lobes show velocities ranging
between $\sim$20 and -20 \kms, respectively \citep{alves15}.  The MeerKAT
bubble appears to be filled with hot coronal X-ray gas
\citep{nakashima13,nakashima19,ponti19} indicating that an energetic outflow
 took place few times 10$^5-10^6$ years ago. These structures are distributed
in the inner 75pc of the Galaxy where high cosmic ray ionization rate
has been inferred, as discussed below. The triggering event is likely to
be enhanced accretion onto the 4$\times10^6$ \msol\, black hole, Sgr A*,
or a burst of star formation activity or both that took place
few times $10^5-10^6$ years ago \citep[e.g.][]{alexander12,zubovas12,wardle14}.



H$^+_3$ absorption measurements toward more than 30 stellar sources indicate that the Galactic center cosmic ray ionization rate ($\zeta$) 
is higher than in diffuse or dense clouds 
in the Galactic disk by two or three orders of magnitudes, respectively 
\citep[e.g.][]{geballe99,oka05,indriolo12,goto14,oka19}. 
Detailed modeling yields $\zeta=2\times10^{-14}$ s$^{-1}$ 
\citep{oka19} and $(1-11)\times10^{-14}$ s$^{-1}$ \citep{lepetit16}. These high values explain three important characteristics of the gas in 
this region.

One, cosmic ray heating explains  the pervasive distribution of warm molecular gas (T$\sim$ 75-200K) compared to the $20-30$K dust temperature throughout the 
Galactic center \citep{price00}, a result of heating by cosmic rays \citep{gusten81,zadeh07b}.

Two, cosmic ray interactions with the ISM 
 produce  a significant fraction of the steady and variable components of  FeI K$_{\alpha}$ line emission at 6.4 keV in the Galactic center \citep{zadeh07a,zadeh13}.

Three, nonthermal bremsstrahlung produced by GeV cosmic ray electrons interacting with neutral gas produces the $\gamma$-ray emission detected by Fermi. Compelling 
evidence for this mechanism is the observed broken power law spectrum of $\gamma$-ray emission which is consistent that seen in the radio spectrum \citep{zadeh13}. In this 
picture, the $\gamma$-ray flux is proportional to the product of gas density and nonthermal radio flux. The average gas density estimated from fitting $\gamma$-ray data is 
similar to that found from H$_3^+$ measurements, less than 100 cm$^{-3}$ \citep{zadeh13,oka19}. This agreement confirms that most of the gas in the inner 2\deg\, by 1\deg\ 
of the Galactic center is filled by warm, diffuse, low density gas clouds and that dense gas has low volume filling factor \citep{oka19}.

Here, we examine the origin of extended coronal gas, nonthermal radio filaments and vertical magnetic fields, all extending  away from the plane and 
arising because of overpressured  cosmic rays in the CMZ driving an outflow.


\section{Cosmic Ray Driven Winds} 

Cosmic rays drive winds in the nuclei of galaxies \citep[][and references therein]{kulsrud71, zweibel17}. They also play a role in feedback, limiting star 
formation and the growth of the central supermassive black holes by transferring their momentum and energy into the surrounding medium. Here we interpret the large scale X-ray and 
synchrotron emission above and below the CMZ as emission arising from a cosmic-ray driven outflow, as drawn schematically in Figure 1 \citep[e.g.][]{breitschwerdt91,everett08,everett10a,ruskowski17}.  
We also explain the origin of warm ionized gas and dust along the eastern and western edges of the 
bubble   as the consequence of coronal gas pushing the 
warm ionized gas mixed in with dust in the direction where the lobes of the bubble  lie, 
l$\sim0.2^\circ$ and $\sim-0.5^\circ$, extending up to a degree away from the Galactic plane. The high energy density of cosmic rays $\ge10^{3}$ eV cm$^{-3}$ in 
the Galactic center suggests global injection of relativistic particles into the CMZ by an 
energetic event such as black hole activity or due to multiple 
supernova explosions. 
The MeerKAT bubble and the bipolar X-ray features are likely relics of this event 
\citep{heywood19}.

Cosmic-ray driven outflow has been used to explain thermal
X-ray and synchrotron emission from the inner few kpc of the Galaxy
\citep{everett10a}. In this picture, the cosmic rays momentum and energy are
mediated by the magnetic field and are transferred to accelerating and heating the
gas. Mass loss from the inner $\sim$2-3 kpc of Galactic disk has been estimated by
fitting the X-ray and synchrotron emission indicating that gas and cosmic ray
pressures launch a wind \citep[see Fig. 5 of][]{everett10a}. The best joint fit to
the soft X-ray and radio synchrotron emission yields a pressure
$\sim5.5\times10^4$ cm$^{-3}$\, K at the base of the wind \citep{everett10a}.
We are considering a similar scenario except that it is on 
the 300 pc scale of the bipolar X-ray
emission and  warm ionized gas in the MeerKAT bubble. The radio bubble  and X-ray chimney features
lie at the base of
a much larger scale structure,  the well-known "Fermi bubbles", where  outflowing gas has been detected
symmetrically on a scale of a few kpc away from the
Galactic plane \citep{su10,yang13}. Recent analysis of Fermi data suggests excess $\gamma-ray$ emission on a scale 
similar to MeerrKAT bubble \citep{heywood19,herold19}.  
The X-ray and synchrotron sources are likely  to be scaled-down
version  of an  outburst activity that produced the Fermi bubbles.
\citep{nakashima19,ponti19,heywood19}. 

Here we examine whether the observed bremsstrahlung emission from the coronal gas in chimneys, warm ionized gas peaking in 
the lobe and radio synchrotron emission are consistent with a cosmic-ray driven wind scenario.

We note that both $\gamma$-ray and H$^+_3$ measurements indicate a low average gas density in the CMZ. This may  be the result of a wind driven outflow removing the gas to high altitudes and regulating star formation in the CMZ. 
However, the high value of $\zeta$ suggests that past nuclear activity may have been more likely in removing gas from the disk of the 
Galaxy by inducing a wind rather than an outburst of star formation.

Assuming a 10\% helium to hydrogen ratio by number and full ionization, the pressure of the thermal X-ray plasma is  $P_g / k 
\approx 1.3\times 10^6$\,K\,cm$^{-3}$, the thermal energy density $U_g = \frac{3}{2}P_g \approx 170$\,eV\,cm$^{-3}$, density $\rho\approx 
2\times10^{-25}$\,g\,cm$^{-3}$ and the sound speed around 400\,km\,s$^{-1}$. 
Similarly, thermal RRL emission gives the gas pressure $P_g / k \approx 3.8\times 
10^6$\,K\,cm$^{-3}$ with energy density $U_g \approx 490$\,eV\,cm$^{-3}$.

\newcommand{\persec}{\,s$^{-1}$}
\newcommand{\percc}{\,cm$^{-3}$}
The cosmic-ray pressure $P_c$ in the wind, should be compatible with that needed to maintain the elevated cosmic-ray ionization rate of $\zeta_H = 
10^{-14}$--$10^{-13}$\persec in the CMZ that has been inferred from stellar H$_3^+$ absorption measurements \citep{oka19}.  In the solar neighborhood, the interstellar 
cosmic-ray energy density of 1.8\,eV\percc produces $\zeta_H \approx 3\times10^{-17}$\persec \citep{webber98}, implying that cosmic-ray energy density in the CMZ  
500--5\,000\,eV\percc at the base of the wind.  We expect the value in the wind to be somewhat lower depending on the scale height of the wind, so we adopt $U_c = 
1\,000$\,eV\percc, implying a cosmic-ray pressure $P_c/k = \frac{1}{3} U_c/k \approx 3.9\times10^6$\,K\,cm$^{-3}$.

To estimate the magnetic field strength, we use the observed intensity of the synchrotron emission from the chimney at 4.8 GHz \citep{law08}, i.e.\, $I_\nu \approx 
0.5$\,Jy per 153\,arcsec FWHM beam, and standard, if somewhat uncertain, assumptions about the population of cosmic ray electrons: their energy density is 2\% of the total 
in cosmic-rays, i.e 20\,eV\percc with an $E^{-2}$ power law running between 1\,MeV and 100\,GeV, and that the depth of the source, $L$ is the same as the transverse extent 
on the sky, i.e. 150\,pc. The synchrotron intensity is then computed using $I_\nu = j_\nu L$, with $j_\nu$ the synchrotron emission coefficient for a power-law electron 
spectrum \citep[e.g.][]{rybicki79}, yielding: \begin{equation} I_\nu \approx 0.144 \, \frac{e^3B}{m_ec^2}\, E_\nu\,n(E_\nu) L\,,\end{equation}where $n(E)\,dE$ is the 
number density of cosmic-ray electrons in the energy range $[E,E+dE]$, and\begin{equation} E_\nu = \left(\frac{4\pi m_e c\nu}{3eB}\right)^{\!\!1/2} m_ec^2\end{equation}is 
the characteristic energy of the electrons radiating significantly at frequency $\nu$. With these assumptions we obtain $B=4.3\,\mu$G,   
an order-of-magnitude below the 
equipartition value of 39\,$\mu$G. 
This implies a magnetic energy density $U_B \approx 0.46$\, eV\,cm$^{-3}$, Alfv\'en speed $v_A \approx 27$\,\kms, 
characteristic electron energy $E_\nu\approx$ 8.3\,GeV, and a synchrotron loss time 54\,Myr.

The total pressure $\approx 5.2\times 10^6$\,K\,cm$^{-3}$, is an order of magnitude greater than the upper 
envelope of the range considered in the models of \citet{everett10a}, but, this is compensated by the density in 
the X-ray chimney being 20 times larger than in the winds they consider.  Neglecting the magnetic field, the 
sound speed in the cosmic ray dominated medium is $c_* = ((5P_g + 4P_c)/\rho)^{1/2} \approx 720$\,\kms, 
indicating that the medium is able to expand well away from the Galactic plane, but may not escape the Galaxy. 
The inferred outflow rate is $\dot{M} = 2\pi r^2 \rho c_* \approx 0.075$\,\msol\,yr$^{-1}$, where 
$r\approx75$\,pc is the cylindrical radius of the chimney. 
If the medium  expands away from the plane, the origin of 
outflow driven lobes and Fermi bubbles may be related to each other. 
The inferred pressure of cosmic rays
significantly exceeding  the thermal pressure
is consistent with  cosmic ray driven Galactic wind simulations \citep{salem14,ruskowski17}. 

Bipolar hollow lobe of warm ionized gas surrounds the region where coronal gas is detected.  The temperature and rms density $<n_e^2>^{1/2}$ derived for this gas are 
$T_e\approx 4\,000$\,K and $\approx 10$\percc, respectively \citep{law09,nakashima19}.  However collisional broadening of the higher-frequency recombination lines 
implies much higher densities, $n_e\sim 300-1\,000$\percc, consistent with pressure equilibrium with the coronal gas, and implying volume filling factor $\sim 
10^{-4}$ for this material \citep{law09}.

The bulk of the   material in the lobe or the  northern half of the bubble 
is traced by mid-infrared dust emission from  a $~5\times 10^{6}$\msol neutral medium with mean density $n_H\sim 
300$\percc \citep{bland-hawthorn03} and shell thickness 5\,pc, yielding a surface density $\Sigma_\mathrm{shell}\sim 0.01$\, g/cm$^2$.  If the 
internal bubble of coronal gas is over pressured with respect to the exterior, it is accelerating  the shell at a rate 
$P_\mathrm{tot}/\Sigma_\mathrm{shell}\sim 20 $\,pc\,Myr$^{-2}$.  This is consistent with models in which the expanding bubble has swept up the shell 
within the last 3\,Myr. 

We next consider how the presence of this wind may explain the  nonthermal radio filaments.

\section{Nonthermal  Radio Filaments (NRFs)}

Radio observations have identified a system of magnetized filamentary structures
and large-scale radio bubble within the inner two degrees of the Galactic center
and vertical lobes at the edges of the bubble \citep[e.g.][]{zadeh84,
zadeh86,zadeh04,sofue84,heywood19}. Over the years dozens of long and narrow
linear filaments have been found with aspect ratio of 10-100
\citep{liszt85,gray91,haynes92,larosa04,zadeh86,zadeh04,lang99,law08}. They appear
as isolated, or bundled filaments running parallel to each other. The magnetic
field is aligned along the linear filaments and is mainly directed perpendicular
to the Galactic plane \citep{inoue84,tsuboi95,zadeh97}.

The morphology shows that the lobes and the magnetized linear filaments as two
main components of a coherent structure. The lobes coincide with the
edge-brightened structure of the MeerKAT bubble and almost all of the isolated and
bundled linear filaments are distributed within the bubble. The new MeerKAT
observations suggest strongly that the filaments and the lobes are causally
connected to the origin of the radio bubble. The origin of NRFs that are found
only in the Galactic center can now be understood in the context of their
association with a unique event that produced the MeerKAT bubble
\citep{heywood19}. 

A number of models have been proposed to explain the origin of the NRFs \citep[see 
reviews by][]{bicknell01,ferriere09}. Two models that we focus here use stellar
interactions with the ISM \citep{nicholls95} and the interaction of a Galactic 
wind with a cloud \citep{shore99,dahlburg02,barragan16}. These models address
three key questions related to the filamentation structure: 1) The mechanism
that accelerates particles to GeV energies, 2) why certain filaments are detected
and not others if there is a global mechanism for their production, and 3) why the
magnetic field structure runs mainly perpendicular to the Galactic plane, given
that the magnetic field is azimuthal near the plane of the CMZ
\citep{nishiyama10}.

\subsection{Winds Interacting with Mass-losing Stellar Bubbles}

One of the earliest models proposed that particles are accelerated to high energies at the termination shock of 
stellar wind bubbles \citep{rosner96}. In this scenario, the cosmic rays produced at the termination shock load 
onto pre-existing ordered magnetic fields surrounding the wind bubbles and produce static filamentary structures. 
One key aspect of this model is that the transverse size of the filament matches the size of stellar wind 
bubbles. An extension of this model was proposed by considering the collective winds of W-R and OB stars in a 
dense stellar cluster producing shock waves that accelerate particles to high energies \citep{zadeh03}. Another 
model suggested that the filament of the Snake filament is due to a runaway star intersecting a supernova remnant 
\citep{nicholls95} and leaving behind a trail of synchrotron emission.  This model does not explain numerous 
other NRFs that are not associated with supernova remnants.


An alternative scenario that has been suggested is that the NRFs arise through the interaction of a cloud with the Galactic wind 
\citep{shore99,dahlburg02,barragan16}. Numerical simulations show 
morphologies remarkably similar to the  NRFs. The  filaments are considered to be dynamic structures and the weak magnetic field is 
amplified locally without the need to invoke a large scale ordered magnetic field configuration \citep{shore99}. In this picture, the magnetized wind 
advects the 
cloud and forms  a current sheet in the wake behind the cloud with the filamentary structures resulting from 
 shear-driven nonlinear instability \citep{dahlburg02}.  This picture is similar to the interaction the  magnetized solar wind and a comet. 3-D MHD numerical simulations 
confirm the formation of a highly structured magnetotail  \citep{gregori00}. 
Cloud sizes are generally far larger than the width of the filaments, $\leq$ 5$''$ (0.2pc).


\newcommand{\pyr}{yr$^{-1}$}


Here we consider a hybrid model in which the magnetized filaments are produced at the interaction sites of the wind 
outflowing away from the Galactic plane and embedded stellar wind bubbles.  However, unlike the original model in 
which a strong local ISM magnetic field is assumed to be perpendicular to the Galactic plane (Rosner and Bodo 1996), 
we assume that the weak magnetic field in the wind wraps around the stellar wind bubble and forms a long tail behind 
the star (see Fig. 2).  The outflowing galactic center wind and its magnetic field are directed vertically away from 
the plane, but the horizontal component of the stellar orbital velocity means that magnetic field lines become 
draped over the stellar wind bubble (see Fig. 2).  The combined thermal, cosmic-ray and ram pressure of the wind, 
$P_\mathrm{ext}/k\approx 1.3\times10^7$\,K\percc\ sets the stand-off distance $r_w$ of the stellar wind termination 
shock: \begin{equation}
r_w = \left(\frac{\dot{M}_w v_w}{4\pi  P_\mathrm{ext}}\right)^{\!\!1/2} \approx0.17\,\textrm{pc}
    \label{eqn:rw}
\end{equation}
where we have adopted stellar wind mass loss rates and wind speeds $\dot{M}_w=1\times10^{-6}$\,\msol\pyr and $v_w = 1\,000$\,\kms, respectively, and pushes the shocked stellar wind into a tail stretching in the direction of the wind.

Meanwhile, the wind's magnetic field is draped around the stellar bubble, and is stretched and compressed as the associated plasma is evacuated and flows upwards along the 
field lines, parallel to the tail of shocked stellar wind.  This creates a filament with magnetic pressure equal to or a sizeable fraction of that in the surrounding 
cosmic-ray driven wind, i.e. $B\la 210\,f\,\mu$G, where $f\leq 1$ is the magnetic fraction of the total pressure.

This compressed field is the site of the synchrotron emission observed in the NRFs.  The relativistic electrons responsible for the synchrotron emission may be residual cosmic-ray 
particles from the external wind that have not been expelled during compression of the magnetic field. 
Making the reasonable assumption that the cosmic-ray and magnetic pressures are equal and sum to $P_\mathrm{ext}$ yields intensity 
$48\,\mu$Jy\,arcsec$^{-2}$ at 327\,MHz for a source depth 0.3\,pc, consistent with the observed filaments. .

\begin{figure}
\includegraphics[width=\columnwidth,keepaspectratio=true]{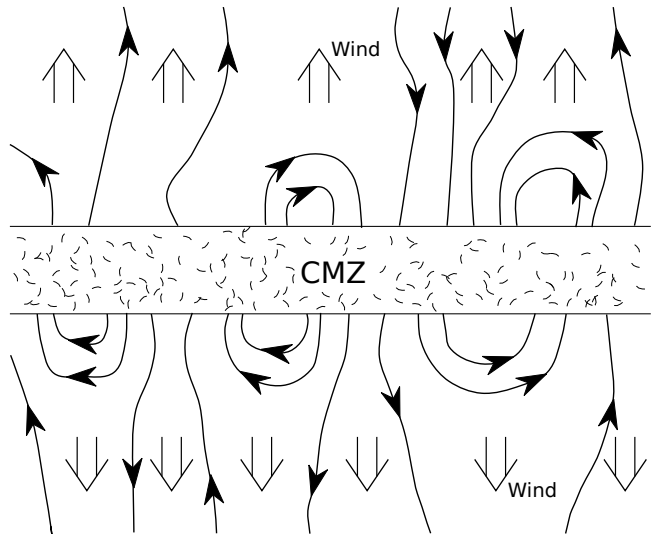}
\caption{\small{A large-scale schematic view of the CMZ azimuthal magnetic field getting distorted to vertical geometry 
by  its interaction with the wind, launched by cosmic rays  through the  Streaming Instability.}}
\label{f:fig1}
\end{figure}

\begin{figure}
\includegraphics[width=\columnwidth,keepaspectratio=true]{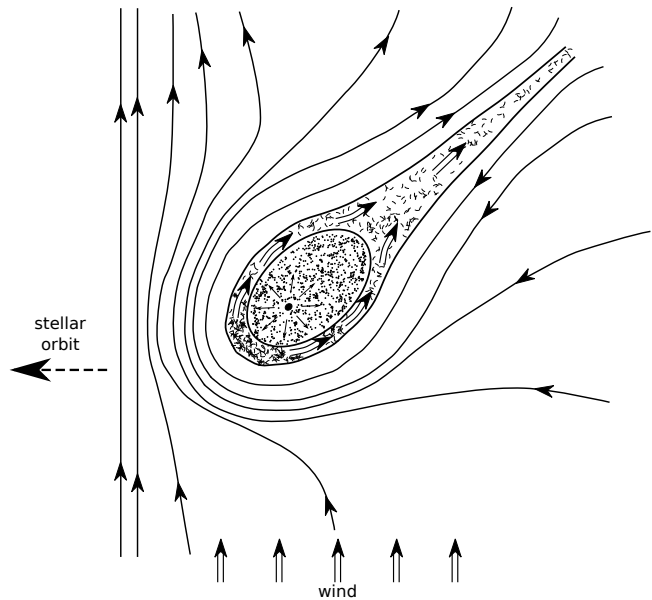}
\caption{\small{A schematic view of the interaction of the magnetized  winds with a stellar bubble moving across 
the field lines}}
\label{f:fig2}
\end{figure}

The stellar wind termination shock is another potential source of relativistic particles.  For example, assuming 
cosmic-ray acceleration efficiency is 0.25 and the proton to electron ratio is 50:1, then the shock luminosity in 
relativistic electrons is $L_e = 0.005 \times \textstyle{\frac{1}{2}} \dot{M_w}v_w^2$, i.e. 0.41\,$\lsol$ for our 
adopted wind parameters.  The electrons are advected away from the shock at the postshock speed, assumed to be 
$v_p = 100$\,\kms (the adopted compression ratio of 10 is greater than for an adiabatic shock because of the 
transfer of energy to cosmic-ray particles), yielding an estimated postshock energy density $L_e/(4\pi r_w^2 v_p) 
\approx 27$\,eV\percc, comparable to the energy density 
 in the magnetic draping scenario outlined in the previous paragraph.
 The origin of the increased 
compression ratio is not just the reduction in adiabatic index from 5/3 towards 4/3 due to the presence of 
relativistic particles (which would increase the compression ratio to 7 at most) but that the shock cools via 
particles escaping upstream \citep{helder12}.
The postshock synchrotron 
brightness is low, however, because the magnetic field there is largely self-generated by the cosmic-ray 
acceleration process and attains $\sim1$\% of the ram pressure i.e. $B\sim 20\, \mu$G \citep{helder12}. However 
this electron population may diffuse into the surrounding region of strong field and contribute significantly to 
the synchrotron emission from that region.

Although most NRFs show linear structure, a number of them  deviate from this geometry. Some  show kinks along their linear structures like the 
Snake  \citep{gray91}, or gently bend, deviating from straight lines perpendicular to the Galactic plane \citep{staguhn98,zadeh16}~and some  run approximately parallel to the 
Galactic plane \citep{lang99,larosa01}. These variations in the morphology of NRFs can be understood in terms  of wind driven outflows having different orientations 
that could vary as a function of time and different orbital velocities of stars. It is possible that the field lines are dragged so much that they get disconnected from 
their parent stellar wind bubble. In this scenario, there is no one-to-one association of filament and stellar wind bubbles. Unlike isolated filaments, a number of 
filaments are bundled and run parallel to each other (e.g., 
\citep{heywood19}.  This network of filaments could be understood if the wind interacts with a cluster of mass-losing stellar wind 
bubbles. The best example of network of NRFs is found near where two well-known Arches and Quintuplet clusters of young stars lie. Alternatively, the network of filaments 
is a 2-D sheet- or cylindrical-like structure made up of individual filaments.


Two puzzling aspects of the nonthermal filaments are the mechanism responsible for 
accelerating particles to relativistic energies and the formation of elongated and 
narrow magnetized filaments, often consisting of multiple filaments running parallel 
to each other.There are no obvious compact sources that can eject relativistic 
particles and illuminate the jet-like appearance of elongated filaments. Here, we 
argued that the cosmic ray driven outflow away from the disk of the Galactic center 
is sweeping across embedded source. In analogy with the interaction of the solar wind 
interacting with the magnetic field of the Earth's atmosphere, we consider that 
outflow with a high cosmic ray pressure is interacting  with the statsphere of an 
embedded mass-losing star. We also discussed the origin of NRFs and why most 
of them run  perpendicular to the Galactic plane. The wind can be produced globally and if the wind is 
related to Fermi Bubble outflows, it may be possible that additional NRFs 
 are expected to be distributed away from the plane in the Fermi bubbles.

In summary, we discussed the consequences of high cosmic ray pressure in the Galactic center region and suggested 
that the coronal gas, the azimuthal magnetic field and the warm ionized gas are driven by an outflow.  We also 
considered that Galactic center nonthermal radio filaments are generated by the interaction sites of wind driven 
Galactic center outflow and embedded stellar wind bubbles.

\section*{Acknowledgments}
This work is partially supported by the grant AST-0807400 from the
the National  Science Foundation.




\bibliographystyle{mnras}


\bsp	
\label{lastpage}
\end{document}